\documentclass[a4paper]{jpconf}

\usepackage{citesort}
\usepackage{graphicx}
\usepackage{xspace}
\usepackage{amssymb,amsmath}
\usepackage{upgreek}
\usepackage{textgreek}

\newcommand{\pT}{\ensuremath{p_{\mathrm{T}}}\xspace}
\newcommand{\pb}{\mbox{Pb--Pb}\xspace}

\newcommand{\ada}{\mbox{A--A}\xspace}

\newcommand{\rs}[1][7 TeV]{\mbox{\ensuremath{\sqrt{s}=} #1}\xspace}

\newcommand{\rsnn}[1][2.76 TeV]{\mbox{\ensuremath{\sqrt{s_{\mathrm{NN}}}=} #1}\xspace}
\newcommand{\rsnno}{\mbox{\ensuremath{\sqrt{s_{\mathrm{NN}}}}}\xspace}

\newcommand{\gvc}{\mbox{\rm GeV$\kern-0.15em /\kern-0.12em c$}\xspace}
\newcommand{\gvcc}{\mbox{\rm GeV$\kern-0.15em /\kern-0.12em c^2$}\xspace}
\newcommand{\mvc}{\mbox{\rm MeV$\kern-0.15em /\kern-0.12em c$}\xspace}
\newcommand{\mvcc}{\mbox{\rm MeV$\kern-0.15em /\kern-0.12em c^2$}\xspace}

\newcommand{\pion}{\ensuremath{\uppi}\xspace}
\newcommand{\pix}{\ensuremath{\pion^{\pm}}\xspace}
\newcommand{\pim}{\ensuremath{\pion^{-}}\xspace}
\newcommand{\pip}{\ensuremath{\pion^{+}}\xspace}

\newcommand{\kx}{\ensuremath{\mathrm{K}^{\pm}}\xspace}
\newcommand{\km}{\ensuremath{\mathrm{K}^{-}}\xspace}
\newcommand{\kp}{\ensuremath{\mathrm{K}^{+}}\xspace}

\newcommand{\ks}{\ensuremath{\mathrm{K^{*0}}}\xspace}

\newcommand{\ksm}{\ensuremath{\mathrm{K^{*}(892)^{0}}}\xspace}
\newcommand{\ksbm}{\ensuremath{\mathrm{\overline{K}^{*}(892)^{0}}}\xspace}

\newcommand{\ph}{\ensuremath{\upphi}\xspace}
\newcommand{\phm}{\ensuremath{\ph(1020)}\xspace}

\newcommand{\ksk}{\ensuremath{\ks\kern-0.15em /\kern-0.12em\mathrm{K}}\xspace}
\newcommand{\kskx}{\ensuremath{\ks\kern-0.15em /\kern-0.12em\langle\kx\rangle}\xspace}
\newcommand{\kskm}{\ensuremath{\ks\kern-0.15em /\kern-0.12em\km}\xspace}
\newcommand{\kskp}{\ensuremath{\ks\kern-0.15em /\kern-0.12em\kp}\xspace}

\newcommand{\phipi}{\ensuremath{\ph\kern-0.15em /\kern-0.12em\pion}\xspace}
\newcommand{\phipix}{\ensuremath{\ph\kern-0.15em /\kern-0.12em\langle\pix\rangle}\xspace}
\newcommand{\phipim}{\ensuremath{\ph\kern-0.15em /\kern-0.12em\pim}\xspace}
\newcommand{\phipip}{\ensuremath{\ph\kern-0.15em /\kern-0.12em\pip}\xspace}

\newcommand{\phik}{\ensuremath{\ph\kern-0.15em /\kern-0.12em\mathrm{K}}\xspace}
\newcommand{\phikx}{\ensuremath{\ph\kern-0.15em /\kern-0.12em\langle\kx\rangle}\xspace}
\newcommand{\phikm}{\ensuremath{\ph\kern-0.15em /\kern-0.12em\km}\xspace}
\newcommand{\phikp}{\ensuremath{\ph\kern-0.15em /\kern-0.12em\kp}\xspace}

\newcommand{\omphi}{\ensuremath{\Omega\kern-0.15em /\kern-0.12em\ph}\xspace}
\newcommand{\omxphi}{\ensuremath{(\Omega^{-}+\bar{\Omega}^{+})\kern-0.15em /\kern-0.12em\ph}\xspace}
\newcommand{\ommphi}{\ensuremath{\Omega^{-}\kern-0.15em /\kern-0.12em \ph}\xspace}
\newcommand{\ompphi}{\ensuremath{\bar{\Omega}^{+}\kern-0.15em /\kern-0.12em \ph}\xspace}

\newcommand{\raa}{\ensuremath{R_{\mathrm{AA}}}\xspace}
\newcommand{\dndy}{\ensuremath{\mathrm{d}N\kern-0.15em /\kern-0.12em\mathrm{d}y}\xspace}

\newcommand{\stpc}{\ensuremath{\sigma_{\mathrm{TPC}}}\xspace}

\begin{document}
\title{Hadronic resonances in heavy-ion collisions at ALICE}

\author{A G Knospe (for the ALICE Collaboration)}

\address{Department of Physics, The University of Texas at Austin, 1 University Station C1600, Austin, TX 78712-0264 USA}

\ead{anders.knospe@cern.ch}

\begin{abstract}
Properties of the hadronic phase of high-energy heavy-ion collisions can be studied by measuring the ratios of hadronic resonance yields to the yields of longer-lived particles.  These ratios can be used to study the strength of re-scattering effects, the chemical freeze-out temperature, and the lifetime between chemical and kinetic freeze-out.  The restoration of chiral symmetry during the early hadronic phase or near the phase transition may lead to shifts in the masses and increases in the widths of hadronic resonances.  The ALICE collaboration has measured the spectra, masses, and widths of the \ksm and \phm resonances in \pb collisions at \rsnn.  These results, including \raa and the ratios of the integrated resonance yields to stable hadron yields, are presented and compared to results from other collision systems and to theoretical predictions.
\end{abstract}

\section{Introduction}

Strongly interacting matter undergoes the transition from a partonic phase to the hadronic phase around a critical temperature of about 160~MeV~\cite{Borsanyi_TC,Aoki_TC2}.  The yields of various particles at chemical freeze-out can be predicted using thermal models~\cite{AndronicQM2011,PBM2011}, but the measurable resonance yields may differ from these predictions due to re-scattering and/or regeneration in the hadronic phase.  Resonances with lifetimes of a few fm/$c$ may decay during the hadronic phase and their decay hadronic daughters may be re-scattered in the medium, reducing the number of resonances that can be reconstructed~\cite{Bliecher_Aichelin}.  Pseudo-elastic scattering of particles (\textit{e.g.}, $\pim\kp\rightarrow\ksm\rightarrow\pim\kp$) before kinetic freeze-out may increase the resonance yield (regeneration)~\cite{Bleicher_Stoecker,Markert_thermal,Vogel_Bleicher}.  The measurable resonance yields are therefore expected to depend upon the chemical freeze-out temperature and the elapsed time between chemical and kinetic freeze-out.  The model described in~\cite{Torrieri_thermal_2001b,Torrieri_thermal_2001b_erratum,Markert_thermal} makes predictions for the ratios of the yields of resonance to stable-particle yields as functions of these parameters, and in these proceedings the measured $\ksm\kern-0.15em /\kern-0.12em\mathrm{K}$ ratio is compared to that model to extract an estimate of the lower limit of the lifetime of the hadronic phase.

Near the phase transition, chiral symmetry is expected to be partially restored~\cite{Petreczky}.  It is predicted that resonances which decay when chiral symmetry is at least partially restored will exhibit mass shifts or width broadening with respect to the vacuum values of those parameters~\cite{Eletsky,Brown_Rho}.  The fraction of resonances with vacuum properties will be increased by regeneration in the later hadronic phase.  Re-scattering and regeneration are predicted~\cite{UrQMD,Markert_thermal} to have the greatest effect on resonance yields at low \pT ($\lesssim2$~\gvc).

In these proceedings, measurements of the \ksm and \ksbm are averaged and these particles are collectively called \ks.  The mass is omitted from the symbols of the \ks and \ph.

\section{Analysis Method}
This analysis uses the ALICE~\cite{ALICE_detector} Inner Tracking System (ITS, for tracking and vertex finding), Time Projection Chamber (TPC, for tracking and particle identification), and VZERO detector (for triggering and centrality estimation).  The \ks (\ph) resonance signals are extracted in multiple centrality bins from about 9~M events.  The $z$ position of the primary vertex is required to be within 10 cm of the center of the ALICE detector.  The resonances are reconstructed through their hadronic decay channels: $\ks\rightarrow\pix\mathrm{K}^{\mp}$ (branching ratio 66\%) and $\ph\rightarrow\km\kp$ (branching ratio 48.9\%)~\cite{PDG}.  Each decay pion (kaon) candidate is required to have TPC energy loss within 2\stpc of the expected value for pions (kaons).

The resonances are identified via invariant-mass reconstruction, with a combinatorial background estimated using event mixing.  The background-subtracted distribution is fit with a polynomial plus a peak function.  The \ks is fit with a Breit-Wigner peak, while the \ph is fit with a Voigtian peak (convolution of a Breit-Wigner function with a Gaussian) to account for detector effects.  The raw yields are corrected for the efficiency~$\times$~acceptance.  This is extracted from simulated \pb collisions (containing 900,000 \ks and 400,000 \ph), with particle production simulated using HIJING~\cite{HIJING} and interactions with the ALICE detector simulated using GEANT3~\cite{GEANT3}.  A separate correction factor is applied to account for the energy-loss cuts (91\% when a 2\stpc PID cut is applied to both decay daughters).  The \ks (\ph) spectra are fit with a L\'{e}vy-Tsallis~\cite{Tsallis} (Boltzmann-Gibbs blast-wave~\cite{BoltzmannGibbsBlastWave}) function so that the yield can be extrapolated for values of \pT outside the measured region.

\begin{figure}
\centering
\begin{minipage}{15pc}
\includegraphics[width=15pc]{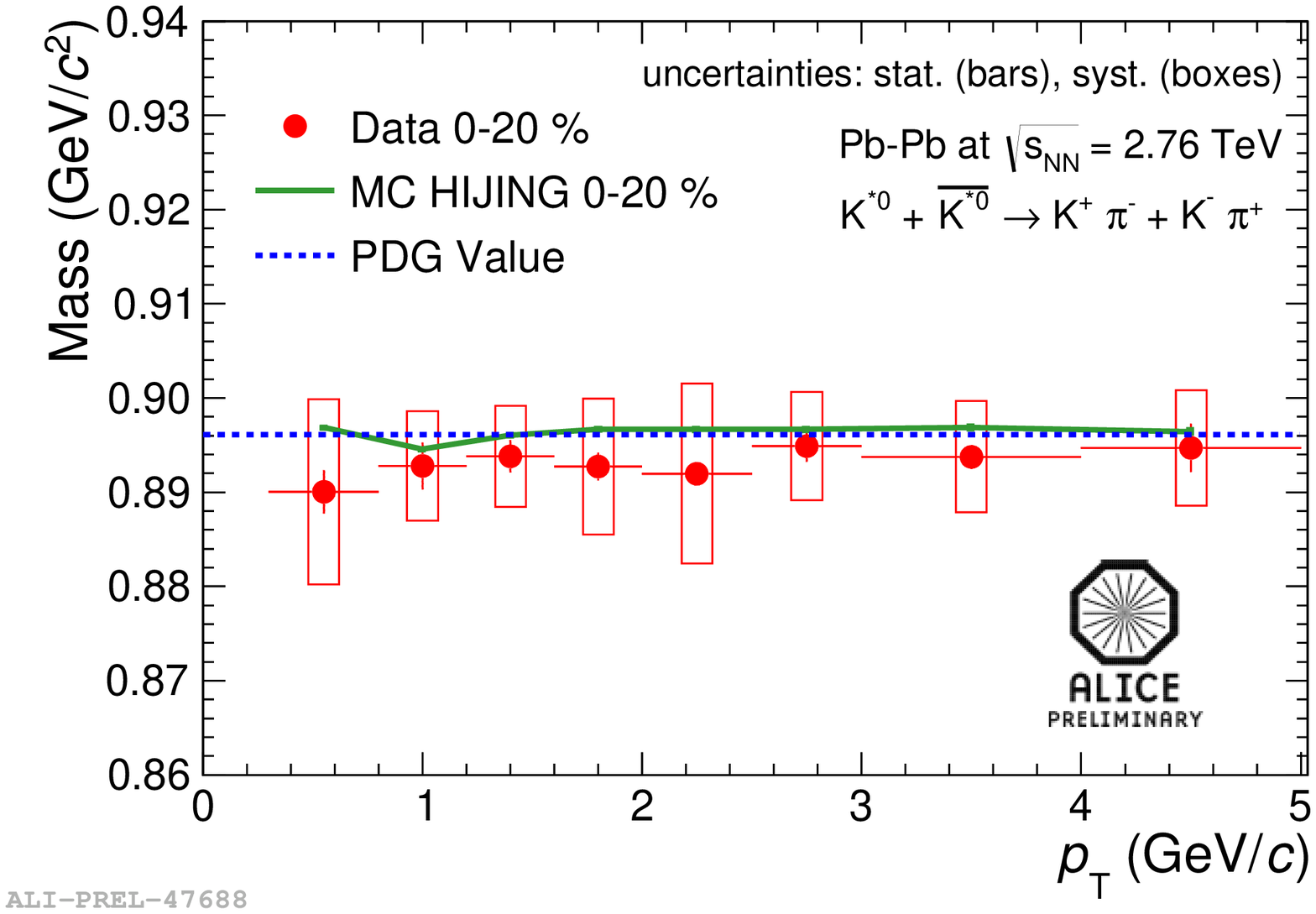}
\end{minipage}
\begin{minipage}{0pc}
\hspace{-13pc}\vspace{-5pc}
(a)
\end{minipage}
\hspace{-1.2pc}
\begin{minipage}{15pc}
\includegraphics[width=15pc]{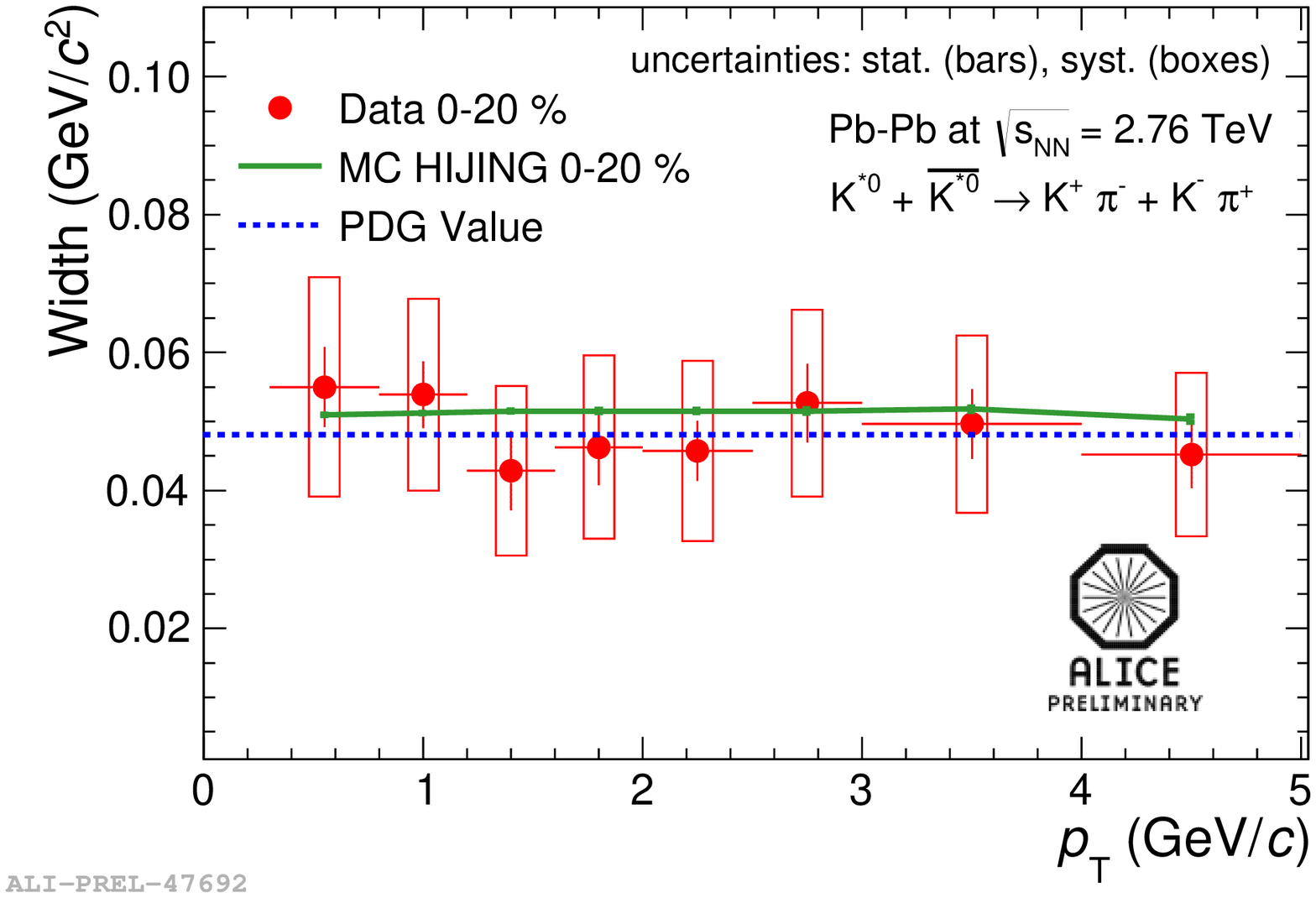}
\end{minipage}
\begin{minipage}{0pc}
\hspace{-13pc}\vspace{-5pc}
(c)
\end{minipage}\\
\begin{minipage}{15pc}
\includegraphics[width=15pc]{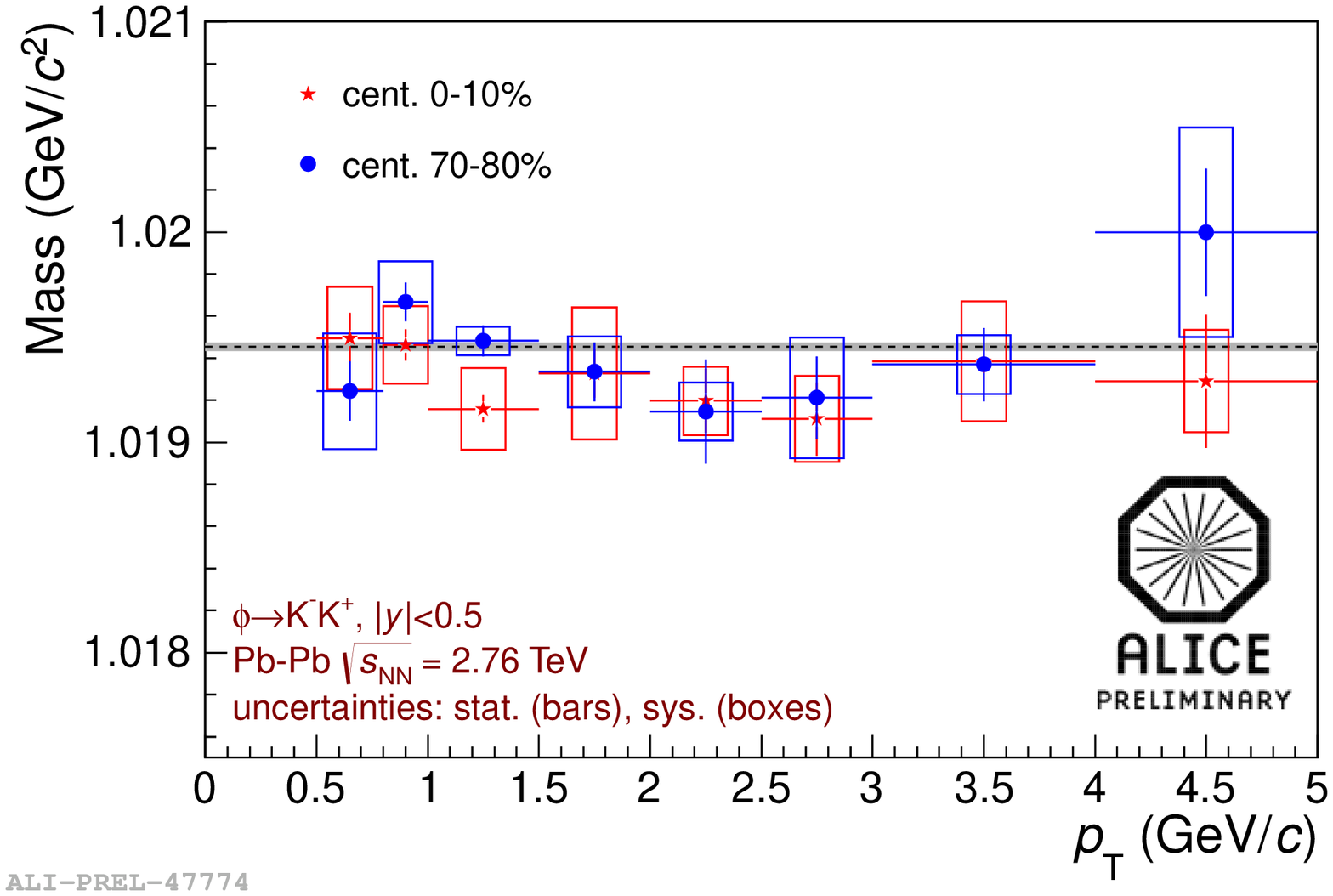}
\end{minipage}
\begin{minipage}{0pc}
\hspace{-13pc}\vspace{-2pc}
(b)
\end{minipage}
\hspace{-1.2pc}
\begin{minipage}{15pc}
\includegraphics[width=15pc]{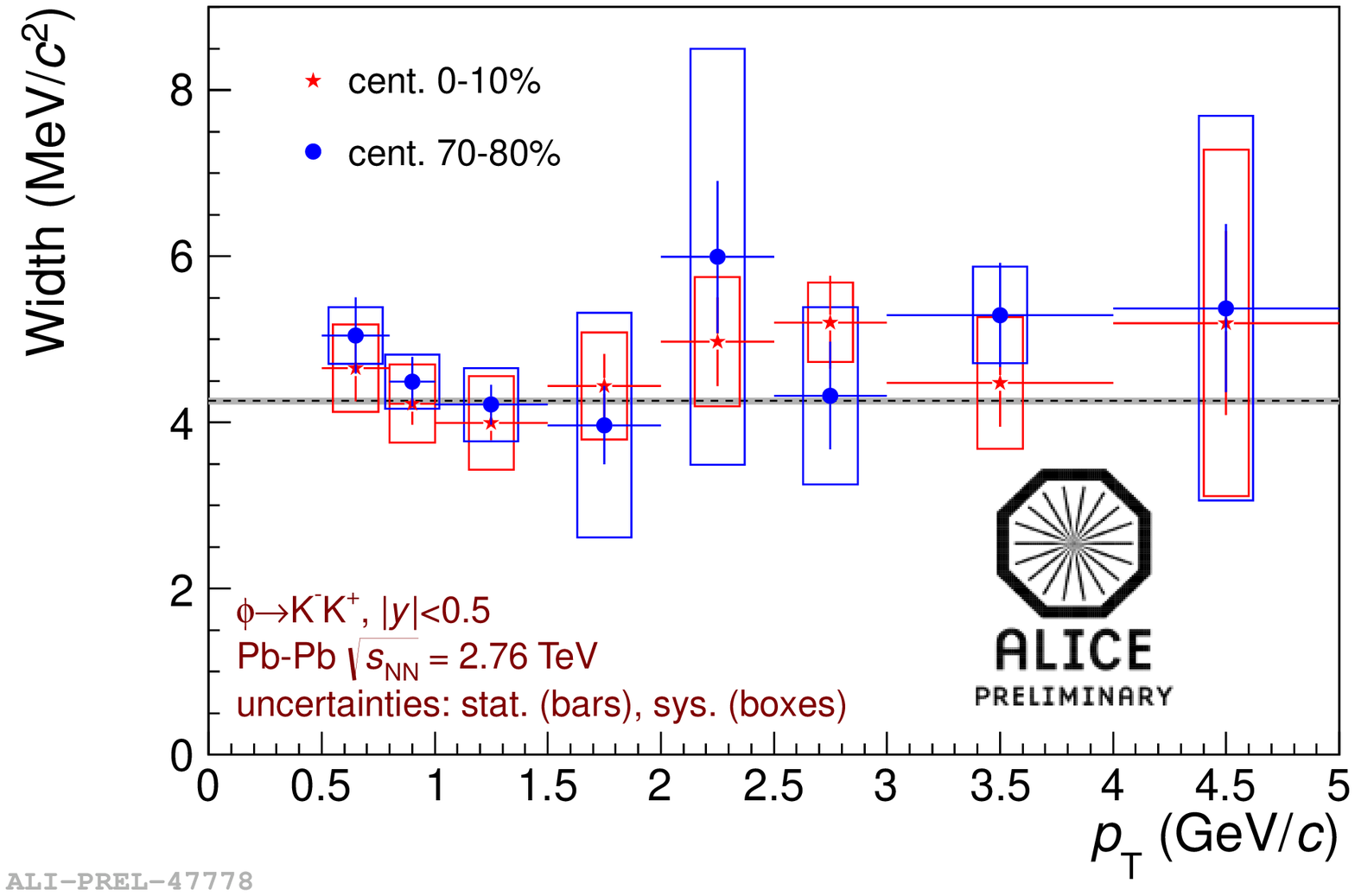}
\end{minipage}
\begin{minipage}{0pc}
\hspace{-13pc}\vspace{-2pc}
(d)
\end{minipage}
\caption{Mass and width of \ks and \ph.} 
\label{fig:mass_width}

\begin{minipage}{15pc}
\includegraphics[width=15pc]{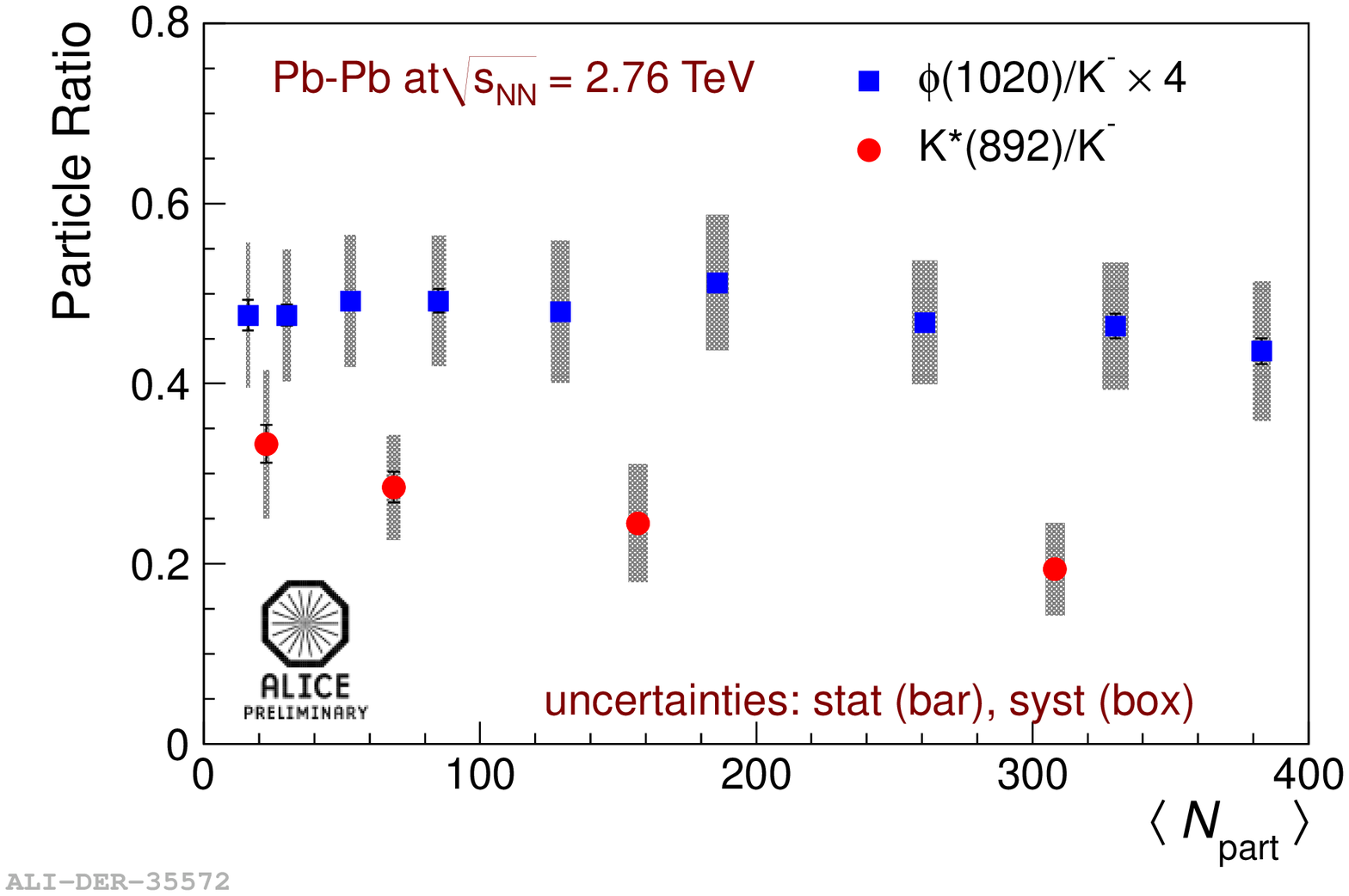}
\end{minipage}
\begin{minipage}{0pc}
\hspace{-13pc}\vspace{6pc}
(a)
\end{minipage}
\begin{minipage}{15pc}
\includegraphics[width=15pc]{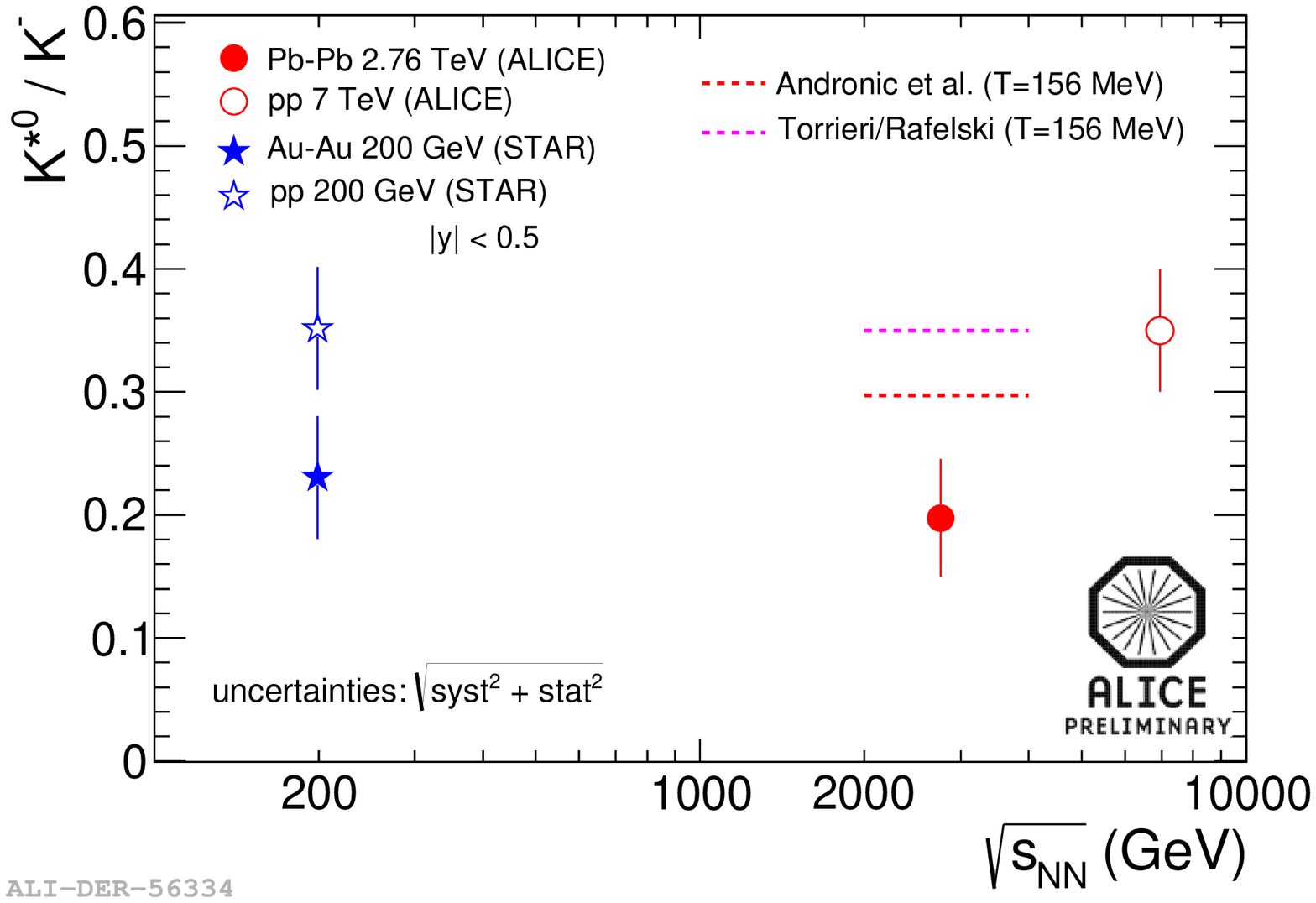}
\end{minipage}
\begin{minipage}{0pc}
\hspace{-13pc}\vspace{-4pc}
(b)
\end{minipage}
\caption{\textbf{(a)} \kskm and \phikm (scaled by 4) ratios as functions of centrality.  \textbf{(b)}  \kskm ratio as a function of \rsnno for different collision systems and theoretical predictions.} 
\label{fig:ratios}

\begin{minipage}{15pc}
\includegraphics[width=15pc]{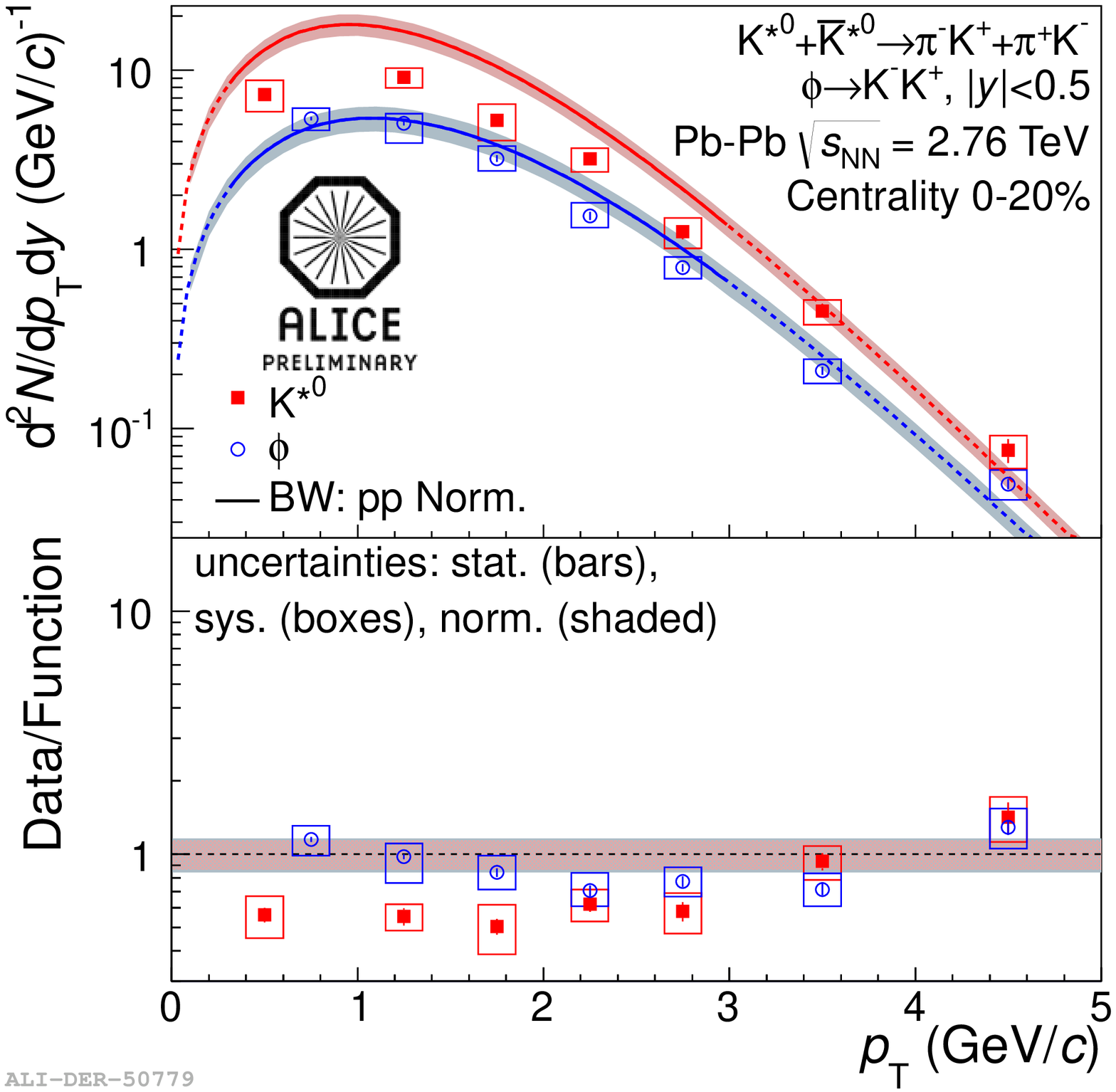}
\end{minipage}
\begin{minipage}{0pc}
\hspace{-13pc}\vspace{-5pc}
(a)
\end{minipage}
\begin{minipage}{15pc}
\includegraphics[width=15pc]{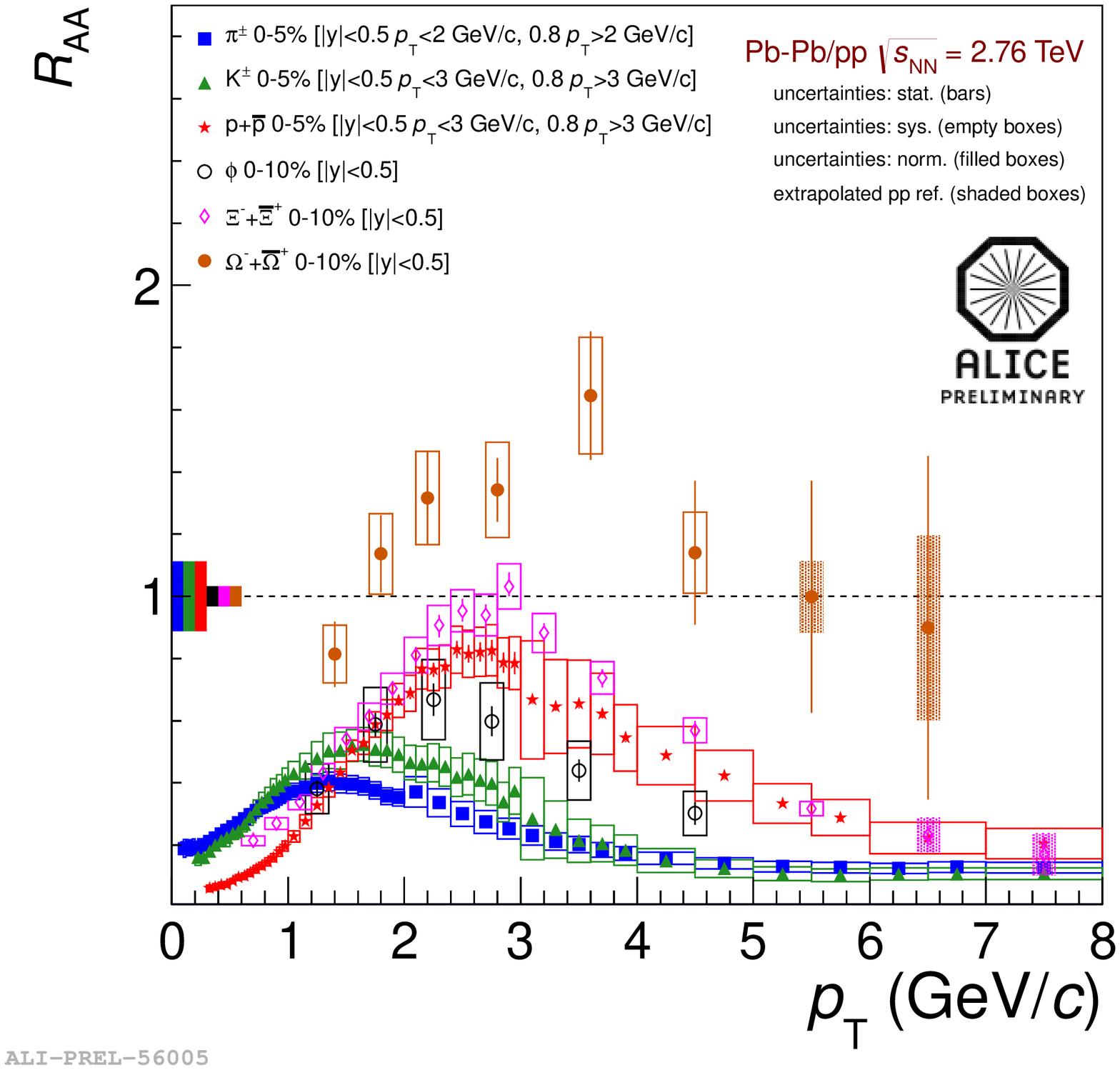}
\end{minipage}
\begin{minipage}{0pc}
\hspace{-13pc}\vspace{-5pc}
(b)
\end{minipage}
\caption{\textbf{(a)} Spectra of \ks and \ph (centrality 0-20\%) with the blast-wave predictions.  The lower panel shows the ratio of the data to the predictions. \textbf{(b)} \raa of various particles.} 
\label{fig:other}
\end{figure}

\section{Results}

The mass and width of the \ks and \ph as functions of \pT in Fig.~\ref{fig:mass_width}.  The measured \ks mass and width are consistent with the values extracted from simulations (the same simulations used to extract the efficiency~$\times$~acceptance).  The measured \ph mass has been corrected to account for the mass shift observed in simulations (\textit{i.e.}, corrected for detector effects).   The \ph mass and width are consistent with the vacuum values~\cite{PDG}.  No centrality dependence is observed in the mass or width for either particle.

The \pT-integrated ratios \kskm and \phikm are shown for different centrality bins in Fig.~\ref{fig:ratios}a (\km yields from~\cite{ALICE_piKp_PbPb_2013}).  The \phikm ratio is independent of centrality, while the \kskm ratio appears to decrease for more central collisions.  This suggests that the \ks is affected by re-scattering, while the \ph (with its longer lifetime) is not.  The \kskm ratio as a function of energy for pp and central \ada collisions~\cite{STAR_Kstar_2011,ALICE_kstar_phi_7TeV} is shown in Fig.~\ref{fig:ratios}b.  The \kskm ratio is independent of energy from RHIC to LHC energies, but a suppression is observed in \ada collisions with respect to pp collisions.  A thermal model fit of ALICE particle yields~\cite{Stachel_SQM2013} (the \ks is excluded from the fit) gives a temperature of 156~MeV and a \ksk ratio of 0.30 (Fig.~\ref{fig:ratios}b, red line), which is ~50\% larger than the measured ratio $0.19\pm0.05$, however this model does not include re-scattering.  The measured \ksk ratio can also be compared to the model described in~\cite{Torrieri_thermal_2001b,Torrieri_thermal_2001b_erratum,Markert_thermal} to estimate properties of the hadronic phase.  This model uses statistical calculations plus re-scattering during the hadronic phase to predict the \ksk ratio (and others) as a function of the chemical freeze-out temperature and the time between chemical and kinetic freeze-out.  If the lifetime of the hadronic phase is assumed to be 0 (\textit{i.e.}, no re-scattering), the measured \ksk ratio would correspond to a chemical freeze-out temperature of $\sim 120$~MeV.  On the other hand, if a chemical freeze-out temperature of 156~MeV is assumed (based on~\cite{Stachel_SQM2013}), the model predicts a $\ksk\approx0.35$ in the case of no re-scattering (Fig.~\ref{fig:ratios}b, magenta line).  However, if a non-zero hadronic lifetime is permitted and a temperature of 156~MeV is again assumed, the measured \ksk ratio gives an estimate of $\approx 1.5$~fm/$c$ as a lower limit on the lifetime of the hadronic phase, according to this model (this is only a lower limit because the model does not include regeneration).

In order to study the \pT dependence of resonance suppression, the measured \ks and \ph spectra for central \pb collisions are compared to predicted spectra based on the blast-wave model~\cite{BoltzmannGibbsBlastWave}.  The blast-wave parameters used to generate the predicted spectra are taken from global fits of the spectra of \pix, \kx, and (anti)protons~\cite{ALICE_piKp_PbPb_2013}.  It should be noted that these fits are not valid for $\pT\gtrsim 3$~\gvc.  The predicted \ks (\ph) spectrum is normalized so that its integral is equal to the K yield in \pb collisions at \rsnn~\cite{ALICE_piKp_PbPb_2013} times the \ksk (\phik) ratio in pp collisions at \rs~\cite{ALICE_kstar_phi_7TeV} (\textit{i.e.}, assuming no re-scattering).  The measured and predicted spectra are shown in Fig.~\ref{fig:other}a, with the ratios of the measured and predicted spectra shown in the lower panel.  The measured \ph spectrum is not suppressed but the measured \ks spectrum is suppressed with respect to the predicted spectrum (normalized as described above).  Furthermore, the \ks suppression is flat $(\approx 0.6)$ for $\pT<3$~\gvc.  In peripheral collisions, neither resonance is observed to be suppressed with respect to the predicted spectra.

The nuclear modification factor \raa for six different particle species is shown in Fig.~\ref{fig:other}b.  For $\pT\lesssim 2.5$~\gvc, the \ph \raa appears to follow \raa of p and $\Xi$, while for mid-to-high \pT, the \ph \raa tends to be between \raa of the mesons (\pion and K) and baryons (p and $\Xi$).  For peripheral collisions, the \ph \raa is consistent with the p and $\Xi$ \raa throughout the measured \pT range and consistent with one for $\pT>1$~\gvc.

\section{Conclusions}

The masses and widths of the \ks and \ph resonances (reconstructed via hadronic decays) are consistent with the vacuum values.  The \ksk ratio decreases for central collisions (which suggests the importance of re-scattering), while the \phik ratio is independent of centrality.  An approximate lower limit of 1.5~fm/$c$ for the lifetime of the hadronic phase has been estimated by comparing the measured \kskm ratio to a theoretical model prediction~\cite{Torrieri_thermal_2001b,Torrieri_thermal_2001b_erratum,Markert_thermal}.  Comparisons of the resonance spectra to predicted spectra based on the blast-wave model indicate that the suppression of the \ks yield is flat for $\pT<3$~\gvc.  The nuclear modification factor \raa of \ph appears to follow \raa of $\Xi$ and $\Omega$ for $\pT\lesssim 2.5$~\gvc, and lies between the \raa of mesons (\pion and K) and \raa of baryons (p and $\Xi$) at high \pT.

\section*{References}
\bibliography{refs}

\providecommand{\newblock}{}
\begin{thebibliography}{10}
\expandafter\ifx\csname url\endcsname\relax
  \def\url#1{{\tt #1}}\fi
\expandafter\ifx\csname urlprefix\endcsname\relax\def\urlprefix{URL }\fi
\providecommand{\eprint}[2][]{\url{#2}}

\bibitem{Borsanyi_TC}
Bors\'{a}nyi$\;$S$\;et\;al$ 2010 {\em J. High Energy Phys.\/} {\bf 2010} No.~9
  1--31

\bibitem{Aoki_TC2}
Aoki$\;$Y$\;et\;al$ 2009 {\em J. High Energy Phys.\/} {\bf 2009} No.~6 1--17

\bibitem{AndronicQM2011}
Andronic A, Braun-Munzinger P and Stachel J 2011 {\em J. Phys.\/} G {\bf 38}
  124081

\bibitem{PBM2011}
Braun-Munzinger P and Stachel J 2011 {\em From Nuclei To Stars: Festschrift in
  Honor of Gerald E Brown\/} ed Lee S (Singapore: World Scientific)
  (\textit{Preprint} \eprint{1101.3167v1})

\bibitem{Bliecher_Aichelin}
Bleicher M and Aichelin J 2002 {\em Phys. Lett.\/} B {\bf 530} 81--7

\bibitem{Bleicher_Stoecker}
Bleicher M and St\"{o}cker H 2004 {\em J. Phys.\/} G {\bf 30} S111--8

\bibitem{Markert_thermal}
Markert C, Torrieri G and Rafelski J 2002 {\em Proc. of PASI 2002\/}
  (\textit{Preprint} \eprint{hep-ph/0206260})

\bibitem{Vogel_Bleicher}
Vogel S and Bleicher M 2005 {\em Proc. of Nucl. Phys. Winter Meeting, Bormio\/}
  (\textit{Preprint} \eprint{nucl-th/0505027v1})

\bibitem{Torrieri_thermal_2001b}
Rafelski J, Letessier J and Torrieri G 2001 {\em Phys. Rev.\/} C {\bf 64}
  054907

\bibitem{Torrieri_thermal_2001b_erratum}
Rafelski J, Letessier J and Torrieri G 2002 {\em Phys. Rev.\/} C {\bf 65}
  069902(E)

\bibitem{Petreczky}
Petreczky P 2007 {\em Nucl. Phys.\/} A {\bf 785} 10--7

\bibitem{Eletsky}
Eletsky V~L, Belkacem M, Ellis P~J and Kapusta J~I 2001 {\em Phys. Rev.\/} C
  {\bf 64} 035202

\bibitem{Brown_Rho}
Brown G~E and Rho M 2002 {\em Rhys. Rep.\/} {\bf 363} 85--171

\bibitem{UrQMD}
Bleicher$\;$M$\;et\;al$ 1999 {\em J. Phys.\/} G {\bf 25} 1859--96

\bibitem{ALICE_detector}
Aamodt$\;$K$\;et\;al$ (ALICE Collaboration) 2008 {\em J. Inst.\/} {\bf 3}
  No.~S08002 i--245

\bibitem{PDG}
Beringer$\;$J$\;et\;al$ (Particle Data Group) 2012 {\em Phys. Rev.\/} D {\bf
  86} 010001

\bibitem{HIJING}
Wang$\;$X$-$N and Gyulassy M 1991 {\em Phys. Rev.\/} D {\bf 44} 3501

\bibitem{GEANT3}
Brun R, Carminati F and Giani S 1994 {\em CERN-W5013\/}

\bibitem{Tsallis}
Tsallis C 1988 {\em J. Stat. Phys.\/} {\bf 52} 479

\bibitem{BoltzmannGibbsBlastWave}
Schnedermann E, Sollfrank J and Heinz U 1993 {\em Phys. Rev.\/} C {\bf 48}
  2462--75

\bibitem{ALICE_piKp_PbPb_2013}
Abelev$\;$B$\;et\;al$ 2013 {\em CERN-PH-EP-2013-019\/} (\textit{Preprint}
  \eprint{arXiv:1303.0737v2})

\bibitem{STAR_Kstar_2011}
Aggarwal$\;$M$\;$M$\;et\;al$ (STAR Collaboration) 2011 {\em Phys. Rev. C\/}
  {\bf 84} 034909

\bibitem{ALICE_kstar_phi_7TeV}
Abelev$\;$B$\;et\;al$ (ALICE Collaboration) 2012 {\em Eur. Phys. J.\/} C {\bf
  72} 2183

\bibitem{Stachel_SQM2013}
Stachel J (for the ALICE Collaboration) {\em these proceedings\/}

\end{thebibliography}

\end{document}